\documentclass[aps,prl,twocolumn,groupedaddress]{revtex4}
\pdfoutput=1

\newcommand{\Vbias}{V_\mathrm{B}}

\newcommand{\df}{\Delta \omega}

\newcommand{\Ec}{E_\mathrm{C}}
\newcommand{\kBol}{k_\mathrm{B}}
\newcommand{\Us}{\mu_\mathrm{2DEG}}
\newcommand{\Uqd}{\mu_\mathrm{QD}}

\usepackage{amsmath}

\newcommand{\bra}[1]{ \langle #1 | }
\newcommand{\ket}[1]{ | #1 \rangle}

\def \Nvoltage {{\cal N}_{V}}
\def \nShell {{n_\text{shell}}}
\def \Dn {D_n}
\def \Dnn {D_{n+1}}
\def \Gm {\Gamma_-}
\def \Gp {\Gamma_+}

\usepackage{graphicx}

\begin{document}

\title{Energy Levels of Few Electron Quantum Dots Imaged and Characterized By Atomic Force Microscopy} 


\author{Lynda Cockins}
\author{Yoichi Miyahara}
\email[Corresponding author: ]{miyahara@physics.mcgill.ca}
\author{Steven D. Bennett}

\author{Aashish A. Clerk}

\author{Peter Grutter}
\email[Corresponding author: ]{grutter@physics.mcgill.ca}
\affiliation{Department of Physics, McGill University,
 3600 rue University,  Montreal, Quebec, H3A 2T8, Canada}

\author{Sergei Studenikin}
\author{Philip Poole}
\author{Andrew Sachrajda}
\affiliation{Institute for Microstructural Sciences,
   National Research Council of Canada, Ottawa, Ontario K1A 0R6, Canada}

\date{\today}

\begin{abstract}
Strong confinement of charges in few electron systems such as in atoms, molecules and quantum dots leads to a spectrum of discrete energy levels that are often shared by several degenerate quantum states.
Since the electronic structure is key to understanding their chemical properties, methods that probe these energy levels in situ are important.
We show how electrostatic force detection using atomic
force microscopy reveals the electronic structure of individual and coupled self-assembled quantum dots. 
An electron addition spectrum in the Coulomb blockade regime, resulting from a change in cantilever resonance frequency and dissipation during tunneling events, shows one by one electron charging of a dot. 
The spectra show clear level degeneracies in isolated quantum dots, supported by the first observation of predicted temperature-dependent shifts of Coulomb blockade peaks.
Further, by scanning the surface we observe that several quantum dots may reside on what topologically appears to be just one. 
These images of grouped weakly and strongly coupled dots allow us to estimate their relative coupling strengths.
\end{abstract}

\maketitle

The ability to confine single charges at discrete energy levels makes 
semiconductor quantum dots (QD) promising candidates
as a platform for quantum computation 
\cite{Loss98,Tanamoto00}
and single photon sources \cite{Dalacu09}.
Tremendous progress has been made in 
not only understanding the properties of single electrons in QDs 
but also in controlling their quantum states which is an essential prerequisite 
for quantum computation \cite{Hanson07}.
Single electron transport measurements have been the main experimental technique 
for investigating electron tunneling into quantum dots \cite{Kouwenhoven01}. 
Charge sensing techniques using \textit{built-in} charge sensors, 
such as quantum point contacts \cite{Field93}, complement transport measurements 
as lower electron tunneling rates can be monitored 
with even real-time detection being possible \cite{Lu2003}.
It is instrumentally challenging to study self-assembled QDs via
conventional transport and charge sensing methods due to the 
difficulty in attaching electrodes.
Although progress is being made
\cite{Ota04, Jung05-2,Igarashi07,Vdovin07,Amaha08} 
these techniques have very small yield and therefore make it difficult
to assess variation in QD electronic properties.
Compared to typical QDs studied via transport measurements, 
in particular lithographically defined QDs, 
self-assembled QDs can be fabricated to have smaller sizes, 
stronger confinement potentials and a more scalable fabrication process,
all of which make them attractive for practical applications.

In this paper, we focus on an alternative technique for studying QDs
that is better suited for self-assembled QDs: charge sensing by 
atomic force microscopy (AFM).
Charge sensing by AFM
is a convenient method to study the electronic structure of QDs
as nano-electrodes are not required and large numbers of
QDs can be investigated in one experiment.
Termed single-electron electrostatic force microscopy ($e$-EFM),
this technique relies on the high force-sensitivity of AFM to detect the electrostatic force 
resulting from single electrons tunneling into and out of the QD.
It was first demonstrated on QDs formed in carbon nanotubes 
\cite{Woodside02, Zhu05}, 
and later applied to self-assembled QDs \cite{Stomp05,Dana05} 
and also to gold nanoparticles \cite{Azuma06, Zhu08}.
We focus on epitaxially grown self-assembled InAs/InP QDs 
in the few-electron regime.
Using a dissipation model, we find compelling evidence for the 
existence of electronic degeneracies (i.e. shell structure)
by measuring an effective temperature-dependent level repulsion; 
although predicted for conductance measurements in 1991\cite{Beenakker91},
we believe this to be the first observation of this effect.
Further, we use the model to quantitatively extract
various properties of both \emph{individual} QDs, 
such as the tunneling rates and charging energy
and \emph{coupled} QDs, such as the strength of coupling.


\begin{figure}
\includegraphics [width = 86mm] {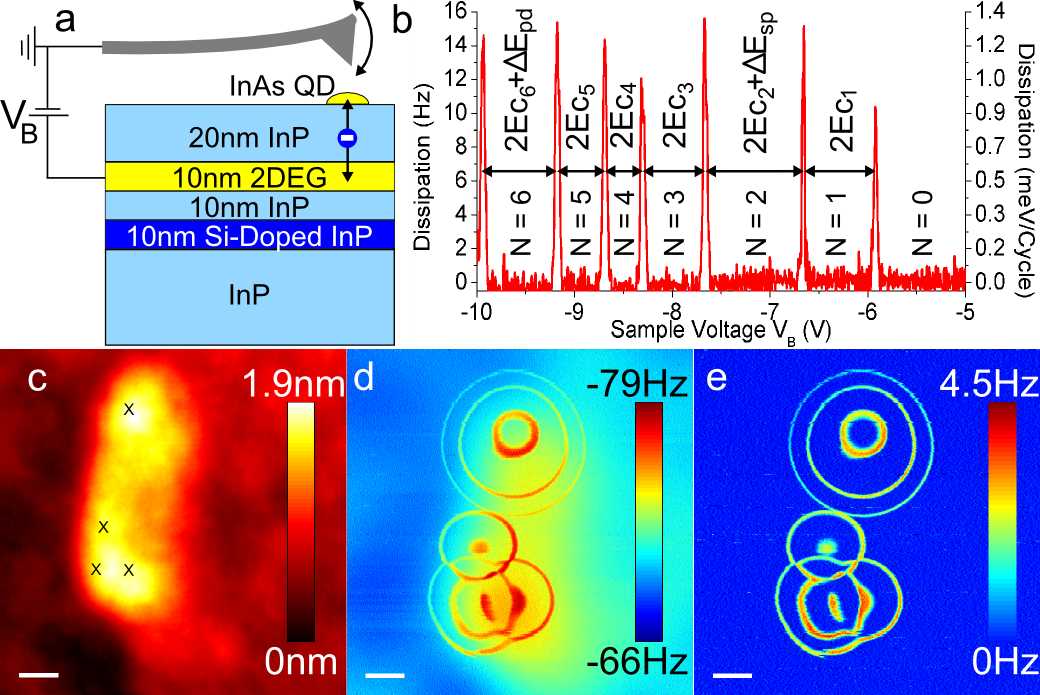}
\caption{\label{fig:mainQD}
{\bf $e$-EFM $\Vbias$-spectra and images.}
{\bf a}, Schematic of the oscillating cantilever with pyramidal tip 
pushing electrons on and off the QD 
when the mean bias voltage is just enough to lift the Coulomb blockade.
{\bf b}, $\gamma$-$\Vbias$ spectra taken at 4.5~K over upper QD shown in {\bf(e)}.
Peaks in the spectra are always separated by the charging energy, 
but shells are additionally separated by $\Delta E$.
After passing each peak from right to left, the number of electrons, $N$, 
in the QD increases by one, with
the $N=7$ state after the leftmost peak.
The energy difference between the first two peaks is 31~meV 
and the difference between peak 2 and 3 is 42~meV, 
so that if $2\Ec$ between peak 2 and 3 is assumed to be 31~meV 
then $\Delta E_{sp}$ = 11~meV.
{\bf c}, Topography of the InAs island with 
the approximate locations of the QDs marked by X's. 
{\bf d} and {\bf e}, The simultaneously recorded frequency shift and dissipation images 
of the structure in {\bf(c)} at 4.5~K taken at $-8$~V.
Scalebar is 20~nm.
}
\end{figure}

We study uncapped self-assembled InAs QDs grown on a 20~nm InP tunnel barrier 
below which a two-dimensional electron gas (2DEG) is formed 
in an In$_{0.53}$Ga$_{0.47}$As quantum well.
A dc-bias voltage, $\Vbias$, is applied to the 2DEG
with respect to the grounded conductive AFM cantilever tip.
Figure~\ref{fig:mainQD}a shows the sample structure and experimental setup.

The AFM cantilever is driven at its mechanical resonance frequency,
$\omega_0/2\pi \sim$166~kHz,
with constant oscillation amplitude \cite{Albrecht91}.
The voltage drop, $\alpha \Vbias$ $(\alpha < 1)$, across the tunnel barrier
between the QD and the 2DEG is only a fraction of $\Vbias$,
with $\alpha= \alpha(x,y,z)$ being a function of the tip position.
The tip-QD gap is tens of nanometers wide
so that tip-QD tunneling is negligible.
We thus have a single-electron box setup:
the electrochemical potential of the 2DEG, $\mu_\mathrm{2DEG}$,
with respect to the QD, $\mu_\mathrm{QD}$, is set by $\alpha\Vbias$
and a negative bias 
increases the number of electrons, $N$,
on the QD
in integer steps whenever the electrochemical potentials are aligned
(called a charge degeneracy point).
Tunneling between 2DEG and QD is suppressed by 
the electrostatic energy cost, $\Ec$, 
of adding or removing an electron to the QD
except near these charge degeneracy points (Coulomb blockade).
The heart of the $e$-EFM technique lies in the fact that oscillations of the AFM cantilever modulate $\alpha$, and hence are equivalent to an effective oscillating gate voltage applied to the QD.  Thus, motion of the cantilever induces a modulation of $N$ which will be slightly out-of-phase with the cantilever's motion (a result of the  finite response time of electrons on the dot).  The electrostatic coupling between QD and cantilever tip implies an electrostatic force proportional to $N$ acting on the tip, the net result being both a frequency shift, $\df$, and additional dissipation, $\gamma$, of the cantilever \cite{Holscher01}.
These effects are maximal at charge degeneracy points as here $N$ can easily change in response to the effective oscillating gate voltage.

Figure~\ref{fig:mainQD}b is an example of $\gamma (\Vbias)$ 
 at 4.5~K 
with the tip positioned over the upper QD imaged in Fig.~\ref{fig:mainQD}d-e.
The $\gamma (\Vbias)$
is equivalent to the energy addition spectra usually
obtained from linear conductance or capacitance spectroscopy measurements \cite{Ashoori92}.
Similar to those measurements, Coulomb blockade peaks in $\gamma$ 
occur near charge degeneracy points of the QD.
The peak positions are suggestive of the addition spectrum of a 2D circular QD with parabolic confinement potential; each peak is separated by twice the capacitive charging energy, $2\Ec$, with a further splitting between peaks 2 and 3 and between peaks 6 and 7 corresponding to the energy difference between shells, $\Delta E$.
This type of shell structure has been previously observed in InAs QDs \cite{Drexler94, Miller97, Ota04, Jung05-2}. 

Figure~\ref{fig:mainQD}c-e shows the topography, $\df$ and $\gamma$ images 
of an elongated InAs island.
The peaks in the $\gamma$-$\Vbias$ spectra radially surround the QD center
so that the ring furthest from the center corresponds to the first electron entering the QD;
the rings themselves are contour lines of constant $\alpha\Vbias$.
Multiple sets of concentric rings appearing in the $\df$ and $\gamma$ images indicate 
multiple QDs in the island.
Such observations would not be as easily identified  
via other experimental means \cite{Zhitenev99}.
The tip-2DEG capacitive force
adds a large background in the $\df$ signal 
which locally varies due to topography
\cite{Stomp05,Cockins07},
making it advantageous to focus on the $\gamma$ for image analysis.

\begin{figure}
\includegraphics [width = 86mm]{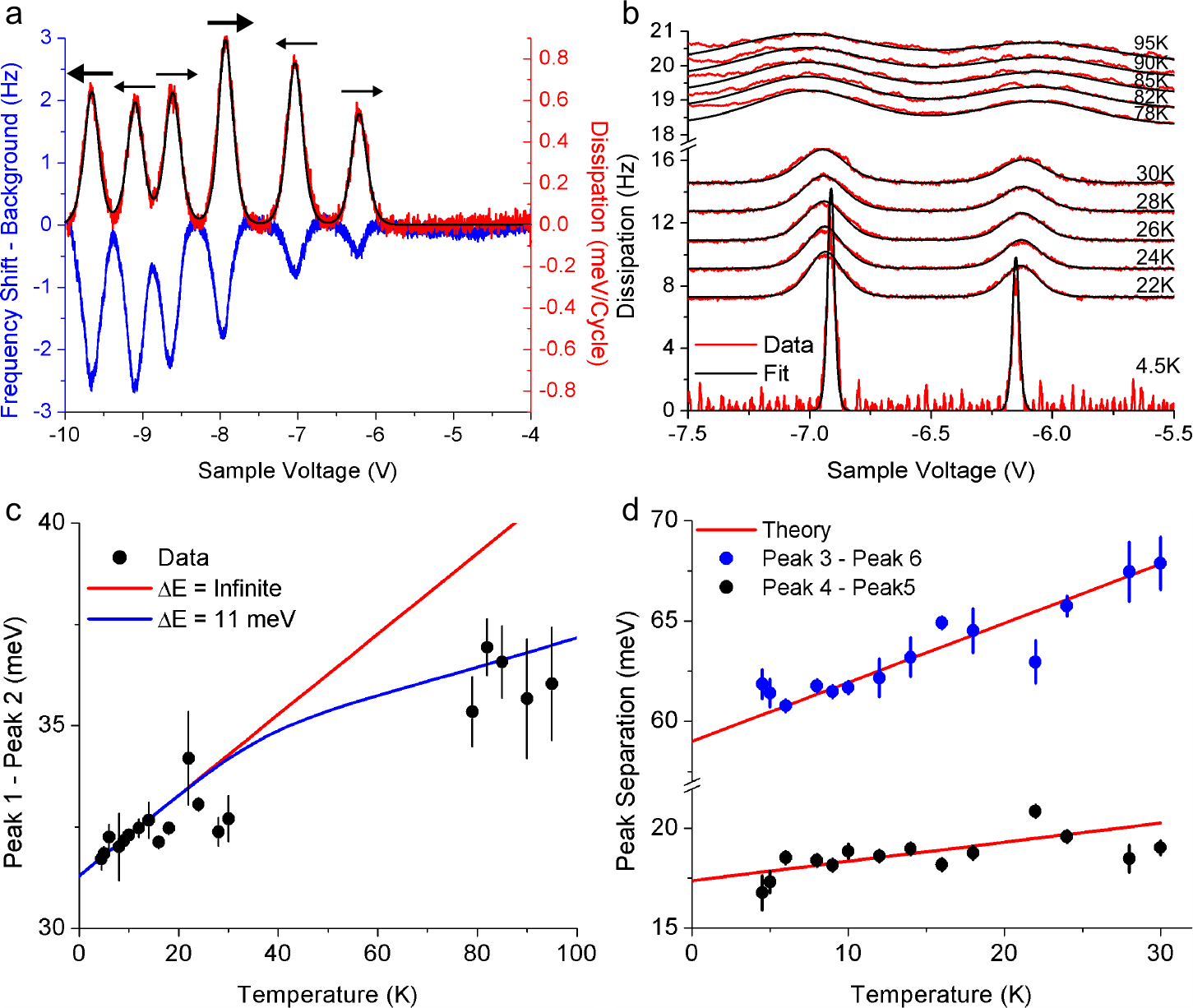}
\caption{\label{fig:spectra}
{\bf Temperature dependent shifts of Coulomb blockade peaks.}
{\bf a}, The dissipation $\gamma$ (red) and frequency shift
$\Delta\omega_{\mathrm{dip}}$ (blue)
measured simultaneously versus bias voltage, $\Vbias$,
over the center of the upper set of concentric rings in Fig.~\ref{fig:mainQD}e.  
Note that the parabolic background due to the capacitive 
force has been subtracted from $\Delta\omega$ 
to show $\df_{\mathrm{dip}}$ (see Methods).  
The lever arm, $\alpha$, is determined by fitting each peak 
to equation (\ref{eq:gammaGeneral}) (black). 
{\bf b}, The first two peaks from the right in $\gamma$ fitted to equation (\ref{eq:gammaGeneral}) 
at different temperatures (offset for clarity). 
{\bf c}, The measured and theoretical separation 
between peaks 1 and 2 as a function of temperature
where the sole fit parameter is the peak separation at zero temperature, $2E_\mathrm{C1}=$31~meV.
A numerical calculation of this separation including
the effects of the empty $p$ shell is also shown (blue).
{\bf d}, The separation between peaks 3 and 6 and peaks 4 and 5.  
Due to thermal broadening of the peaks, 
the positions of these peaks could only be determined up to 30~K. 
The directions and magnitudes of the peak shifts 
as a function of temperature are indicated with arrows 
in {\bf (a)}, with larger arrows indicating greater shifts.}
\end{figure}

Figure~\ref{fig:spectra}a shows $\gamma$ and $\Delta\omega$ 
as a function of $\Vbias$ with the tip positioned
over the center of the upper set of concentric rings 
in Fig.~\ref{fig:mainQD}e at 30~K.  
While the energy addition spectra shown
in Fig.~\ref{fig:mainQD}b and \ref{fig:spectra}a are consistent 
with the expected shell structure for a 2D circular QD
with parabolic confinement potential, 
we obtain much stronger evidence of the shell structure from the 
temperature dependence of the peak positions.
Theoretically, a temperature-dependent shift 
of Coulomb blockade peaks is expected 
whenever one has degenerate single particle levels,
as predicted for the conductance peaks of a spin degenerate level \cite{Beenakker91}.  
The result is an effective temperature-dependent energy level repulsion: 
the peaks in each shell move apart as temperature is increased.
Furthermore, our theoretical analysis suggests that
this effect should be enhanced in the tunneling-induced dissipation 
compared to the conductance due to an increased sensitivity to degeneracy.
The same theory also leads to asymmetric lineshapes
in the Coulomb blockade peaks, also enhanced compared to the conductance, 
but this is predicted to be small and could not be resolved 
in the present experiment.
However, we were able to measure the temperature-dependent shifts of the peaks 
and find excellent agreement with theory.
We are unaware of any experiments
where these effects have been observed.

We model the dissipation on the cantilever using linear response 
and a master equation describing the charge state of the QD 
in the regime of weak coupling \cite{Beenakker91,Clerk05}.
Details of the approach are provided in the Supplementary Information.
Near a charge degeneracy point between two nondegenerate single particle levels, 
the dissipation is \cite{Zhu08, BrinkThesis}
\begin{equation}
  \label{eq:gamma}
  \gamma = \frac{\omega_0^2 A^2 \Gamma}{k_0 \kBol T } 
  	\frac{1}{\omega^2 + \Gamma^2}f (1-f ),
\end{equation}
where $\omega_0$ and $k_0$ are 
the intrinsic cantilever oscillation frequency and spring constant,
$\omega = \omega_0 + \df$ is the measured resonance frequency due to forces on the cantilever,
$\Gamma$ is the 2DEG-QD tunneling rate, and
$f = 1/(1+\exp(E/k_BT))$ is the Fermi function evaluated 
at $E=\Uqd - \Us= e\alpha(\Vbias + V_0)$
($\Vbias = - V_0$ is the point of charge degeneracy).
Lastly, $A = - 2 \Ec \frac{\Vbias}{e} \left( 1-\alpha \right) \frac{\partial C_\mathrm{tip}}{\partial z}$ 
is the sensitivity of the potential on the QD to the cantilever motion 
and $C_\mathrm{tip}$ is the tip-QD capacitance.
We stress that equation (\ref{eq:gamma}) applies to each dissipation peak independently: 
$\Gamma$ and $A$ are obtained separately for each peak from the data with no assumption of constant $\Ec$. 

Equation (\ref{eq:gamma}) only takes into 
account single nondegenerate levels.
More generally, 
suppose we have a state with $N+1$ electrons, with $n_\mathrm{shell} + 1$ in the valence shell.  If this shell has a degeneracy $\nu$, then near the charge degeneracy point between this state and the state with $N$ QD electrons, the dissipation is:
\begin{equation}
\label{eq:gammaGeneral}
	\gamma(\Vbias) = \frac{\omega_0^2 A^2\Gamma}{k_0 \kBol T} 
	\frac{(n_\mathrm{shell}+1)(\nu-n_\mathrm{shell})}{\omega^2 + (\phi \Gamma)^2} 
	\frac{f(1-f)}{\phi},
\end{equation}
where
\begin{equation}
\label{eq:phi} 
	\phi = (\nu- n_\mathrm{shell}) f + ( n_\mathrm{shell} + 1) (1-f),
\end{equation}
and the tunneling rate $\Gamma$ is assumed
to be equal for each degenerate single particle level
within a shell for a given peak
\cite{StevesFootnote}.
Note that $n_\mathrm{shell}$ is the 
number of electrons occupying the given shell 
and not the total number of electrons on the dot, $N$,
and that because of the factor $\phi$, $\gamma(\Vbias)$ 
is no longer symmetric about its maximum.
The different coefficients of $f$ and $1-f$ in $\phi$ reflect 
the asymmetry between electron addition and removal processes caused by degeneracy, 
and $\phi = 1$ corresponds to a nondgenerate level 
for which equation (\ref{eq:gammaGeneral}) reduces to equation (\ref{eq:gamma}).
This asymmetry in equation (\ref{eq:gammaGeneral}) causes each peak in $\gamma (\Vbias)$
to be shifted in energy by an amount proportional to temperature.
By fitting $\gamma (\Vbias)$  (e.g. Fig.~\ref{fig:spectra}a) to equation (\ref{eq:gammaGeneral})
we extract $\alpha$, allowing us to convert the $\Vbias$ axis into energy.
This is done for all of the thermally limited peaks, yielding 
$\alpha = 0.036 \pm 0.003$.
Figure~\ref{fig:spectra}b shows $\gamma (\Vbias)$ at different temperatures
together with the fitted curves.

The role of $\phi$ is further elucidated by the relation
\begin{equation}
\label{eq:responseTime}
	\phi\Gamma = - 2 \omega_0 \frac{\Delta \omega_\mathrm{dip}}{\gamma},
\end{equation}
where $\Delta \omega_\mathrm{dip}$ is the size of the frequency shift 
dip due to the single-electron tunneling.
The ratio in equation (\ref{eq:responseTime}) defines an inverse timescale 
set by the relative in-phase and out-of-phase parts of the electrostatic force;
this is simply $\Gamma$ for a nondegenerate QD \cite{Zhu08,BrinkThesis},
but modified by degeneracy through the factor $\phi$.
Using equations (\ref{eq:gammaGeneral}) and (\ref{eq:responseTime}) 
and the measured values of $\gamma$ and $\Delta\omega_\mathrm{dip}$, 
we calculate the tunneling rates at the maxima of dissipation peaks 1-6, 
obtaining $\Gamma/2\pi = $ 70, 90, 160, 180, 230 and 330 kHz.
As expected, $\Gamma$ increases with increasing $\Vbias$
as the height of the potential barrier between the 2DEG and the QD is reduced.

After extracting the tunneling rates,
we fit each dissipation peak using equation (\ref{eq:gammaGeneral})
and measure the spacing between peaks
as functions of temperature from 4.5-30 K and from 78-95 K.
We focus on the relative shifts between peaks 
as these are less sensitive to slight offsets in peak positions 
due to small changes in the tip-QD distance.
The size and direction of each peak shift is different
(see Fig.~\ref{fig:spectra}a), in a manner that is 
completely captured by our model:
the two peaks in the $s$ shell shift apart, as do the four peaks in the $p$ shell.
The measured relative peak shifts of repelling pairs are shown 
in Fig.~\ref{fig:spectra}c-d 
and compared to the theoretical shift
from equation (\ref{eq:gammaGeneral}) with a single fit parameter, $\Ec$.
In addition,
we expect that multiple shells, not just the valence shell,
should play a role at high
temperatures where $\kBol T \ll \Delta E$ is not satisfied.  
For the relative shift of peaks 1 and 2 (Fig.~\ref{fig:spectra}c) we plot
a numerical calculation accounting for the possible
occupation of the $p$ shell, showing that the
high temperature correction agrees well with the
data.  
Figure 2d shows that the relative shifts between
peaks 3 and 6 and between 4 and 5 are well
described by equation (\ref{eq:gammaGeneral}) up to 30 K.
Finally we note that there is an overall shift of the $s$ and $p$ shells
toward each other that we believe is a consequence of 
strong repulsion of the $p$ shell by the $d$ shell, 
predicted to be 6-fold degenerate.

We performed several checks to support our conclusion that
the peak shifts result from degeneracy. 
First, the predicted shifts are unchanged if the relevant electronic levels
are not perfectly degenerate, but rather split by an amount smaller than $\kBol T$; 
thus, the effect only requires approximate shell degeneracy as is the reality in imperfect QDs.
Second, if the level splitting is larger than $\kBol T$
we expect no temperature dependence of the peak shifts.
Based on the observed peak shifts down to 4.5 K 
we thus conclude that the level splitting is smaller than this temperature, 
corresponding to roughly 0.4 meV.
Lastly, the same degeneracy theory leads to small, but measurable, shifts 
between the dissipation peak and the frequency peak
corresponding to the same charge degeneracy point.
This is visible for the third peak (from the right) in Fig.~\ref{fig:spectra}a
in which the $\gamma$ and $\df$ peaks do not exactly line up,
with the measured shifts compared to theory in Fig.~S2.
These dissipation-frequency shifts strongly support our model 
and rule out the alternative of a temperature-dependent
renormalization of $\Ec$ or $\Delta E$.


\begin{figure}
\includegraphics [width = 86mm] {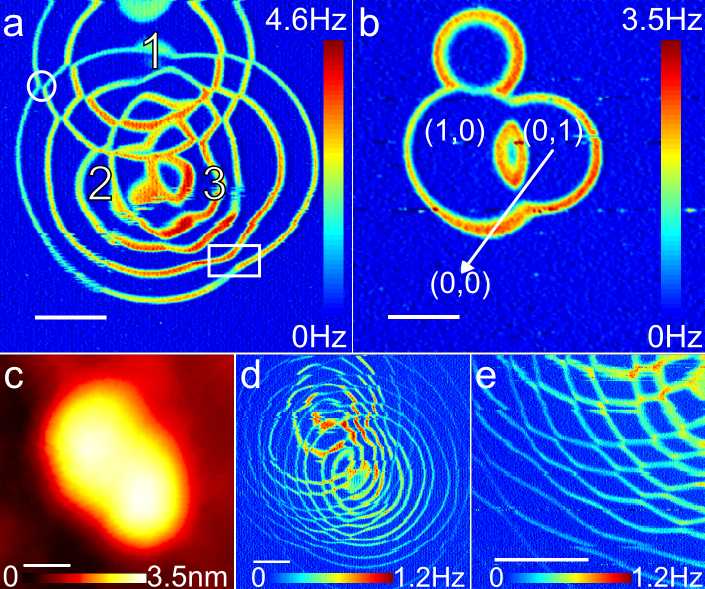}
\caption{\label{fig:doubleQD}
{\bf Imaging coupled QDs.}
{\bf a}, Dissipation image showing the same 3 QDs as the lower half 
of Fig.~\ref{fig:mainQD}e taken at a larger bias voltage, $\Vbias=-9$~V.
{\bf b}, Dissipation image of the same region as {\bf (a)} taken at $\Vbias=-7.6$~V.  
The 3 QDs are numbered in {\bf (a)} but are easier to identify in {\bf (b)} 
since each QD contains one electron.  
In {\bf (a)} an example of weak coupling
between QD1 and QD2 is circled 
and an example of strong coupling between QD2 and QD3 is boxed.
The possible mechanisms are discussed in the text.  
{\bf c}, Topography of two connected islands.
{\bf d}, Dissipation image taken at $\Vbias=-8.0$~V of the structure {\bf (c)}.
Each structure in {\bf (c)} appears to have an associated QD.
{\bf e}, Dissipation image of region in {\bf (d)} with many anti-crossings. 
Scalebar is 20~nm.
}
\end{figure}

Significant efforts are ongoing towards understanding and controlling 
the properties of coupled QDs, in particular double QDs 
or ``artificial molecules'' \cite{Wiel03}.
The $\gamma$ images that we obtain for double QDs are equivalent to stability diagrams 
which depict the charge state of the double QD system.
This is because of the position dependence 
of the lever arm $\alpha (x,y,z)$ for each QD
which results in two electrochemical potentials, 
$\mu_\mathrm{QD1}$ and $\mu_\mathrm{QD2}$.
Scanning the AFM tip at constant height and $\Vbias$ corresponds
to sweeping the $\mu_\mathrm{QD1}$-$\mu_\mathrm{QD2}$ space 
through changing $\alpha_1$ and $\alpha_2$ 
even though only a single electrode (the tip) is being used.

In a conventional stability diagram, lines of constant electrochemical potential 
for each QD are plotted as a function of two gate voltages.
When the two QDs are coupled, intersection points
are split into two points (triple points), showing avoided-crossings \cite{Wiel03}.
In the $\gamma$ images the avoided-crossings
are observed when the ring radii suddenly change at intersection points.
Figure~\ref{fig:doubleQD}a,b show the same three QDs as in the lower part of
Fig.~\ref{fig:mainQD}e, now imaged at $-9$~V and $-7.6$~V, respectively.  
Such avoided-crossings are highlighted in the circle and box 
in Fig.~\ref{fig:doubleQD}a,
representing an example of weak and strong coupling respectively.

We characterize the coupling strength by comparing the
ratio of the change in ring radius to the separation
between the first two rings ($2\Ec$) for QD2.
This method is only valid when both rings are far enough from the QD center
that the voltage drop between them is approximately linear.
Following this procedure, the coupling of QD2 to QD1 (circle) and QD3 (box) 
can be compared.  
While the change in radius of QD2 is approximately 0.10 $\pm$ 0.01 of $2\Ec$ due to QD1,
it is 0.46 $\pm$ 0.03 due to QD3, indicating a much stronger coupling between QD2 and QD3.
We consider the former to be an example of weak coupling because the triple
points are nearly joined.
This is consistent with a small capacitive coupling between the two dots:  the charging of one dot effectively gates the second dot, causing a sudden change in ring radius.

Conversely, the boxed region in Fig.~\ref{fig:doubleQD}a is an example of
strong coupling since there is a large gap at intersections
as in the triple points of a stability diagram.
In Fig.~\ref{fig:doubleQD}b, the same 3 QDs as in Fig.~\ref{fig:doubleQD}a 
are imaged at smaller $\Vbias$. 
This image allows for a more intuitive explanation of the coupling.  
Consider the diagonal line from the center of QD3 outwards;
initially, the AFM tip is over QD3 in the $(N_{\mathrm{QD2}},N_{\mathrm{QD3}})$=(0,1) state,
but takes a path into QD2 in the (1,0) state.
The ability to go continuously between these states without going through (0,0) or (1,1) necessarily indicates a large capacitive coupling between the dots.  It also indicates evidence for an interesting charge transfer process as no dissipation is observed between circles.  Lack of dissipation implies no change in the total dot charge; either there is a cotunneling process where two electrons simultaneously tunnel to and from the 2DEG, or there is coherent tunneling between the dots.

Figure~\ref{fig:doubleQD}c-e shows another example of coupled QDs at 4.5~K.
The InAs structure (Fig.~\ref{fig:doubleQD}c) contains coupled QDs
as shown in the $\gamma$ image (Fig.~\ref{fig:doubleQD}d).
Figure \ref{fig:doubleQD}e zooms up on the region in Fig.~\ref{fig:doubleQD}d 
showing many avoided crossings.
Within this distance range from the QD centers,
each $\alpha$ is approximately linearly dependent on the tip position
so that scanning the tip more closely resembles sweeping two gate voltages,
resulting in the image resembling a conventional stability diagram.
Figure~\ref{fig:doubleQD} also highlights how advantageous it is to have images 
in addition to the $\gamma$-$\Vbias$ spectra 
as the spectra alone will contain the peaks from nearby QDs 
which can be identified using the images.


\begin{figure}
\includegraphics [width = 86mm] {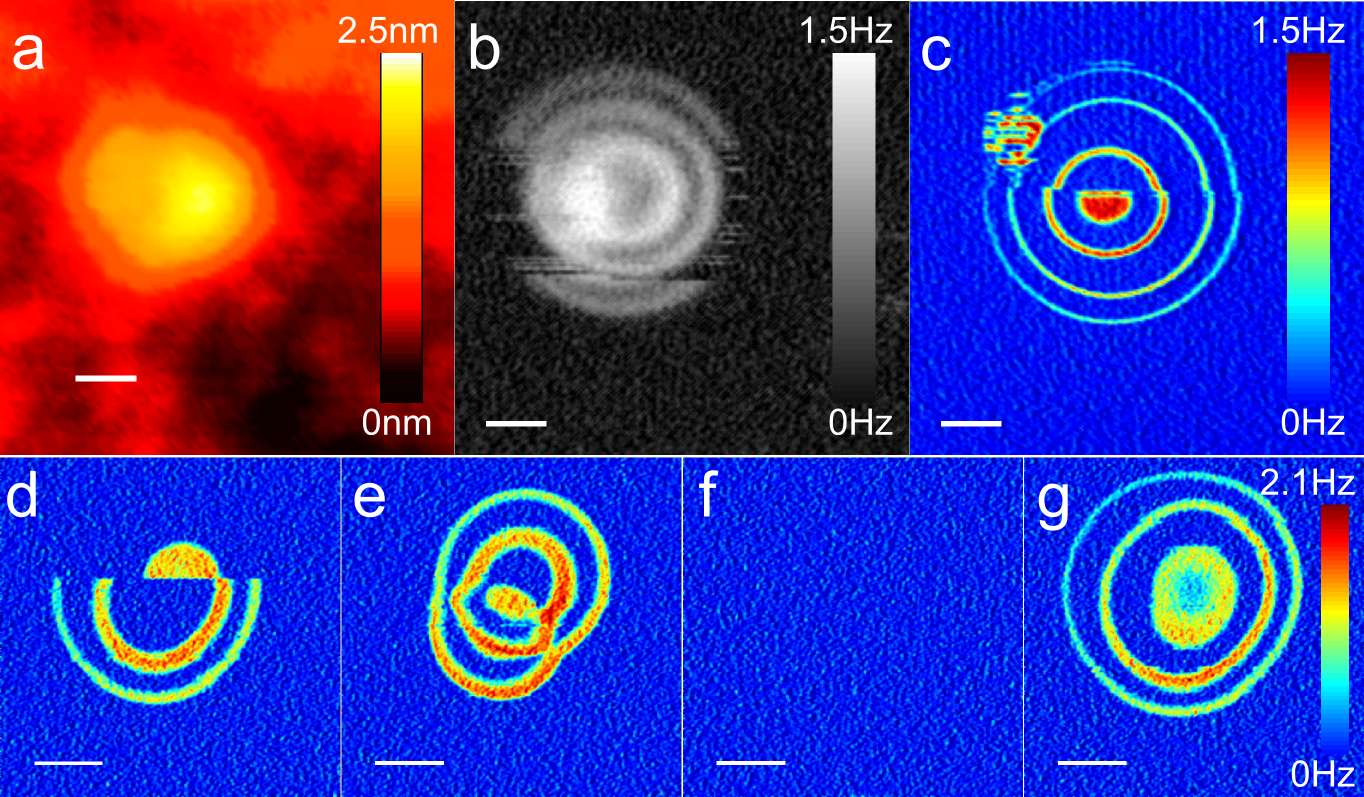}
\caption{\label{fig:switchers}
{\bf Imaging charge noise.}
{\bf a}, Topography of QD.
{\bf b}, Dissipation image of {\bf (a)} at 78~K.
Middle sections of the outermost ring are missing, 
which indicates reconfiguration of nearby charge.
{\bf c}, Dissipation image of {\bf (a)} at 4.5~K.
The slight discontinuity between the upper and lower half of the rings 
indicate that the energy levels of the QD shifted towards $\Us$
in the lower half.
Perhaps the cause of the ``charge noise'' is 
the emerging QD in the upper left corner of {\bf (c)} 
which is in the vicinity of a small protrusion of the structure in {\bf (a)}.
{\bf d}, Dissipation image of a different QD at 4.5~K with $\Vbias=-8.0$~V.
During one horizontal scan line (slow scan direction is upwards), 
the confinement potential abruptly changes.
{\bf e}, Dissipation image of the same region as {\bf (d)} at $\Vbias=-9.0$~V
where coupled QDs are now present.
{\bf f}, Dissipation image taken at $\Vbias=+6.8$~V 
revealing no features as the QD is empty.
{\bf g}, Dissipation image taken immediately after {\bf (f)} at $\Vbias=-8.0$~V. 
The original confinement potential was restored.
Scalebar is 20~nm.  {\bf (d)-(g)} Share the same colorbar.
}
\end{figure}

Yet another advantage of our technique is the ability to spatially resolve
the effects of background charge fluctuations;
for self-assembled QDs, no other technique is capable of doing this.
The AFM images show how the QD confinement potential 
is being influenced by charge reconfiguration.
Figure~\ref{fig:switchers} shows two such cases.
The structure in Fig.~\ref{fig:switchers}a
shows a fluctuation in electron population 
due to nearby fluctuations in the electrostatic background
at both 78~K (Fig.~\ref{fig:switchers}b) and 4.5~K (Fig.~\ref{fig:switchers}c).  
The missing sections of the first ring in Fig.~\ref{fig:switchers}b indicate 
that the number of electrons in the QD is fluctuating by one in this region.  
Interestingly, depending on the scan direction (left to right or right to left) 
over the QD, the missing section may appear.
Similar reconfiguration was observed in the $\gamma$-$\Vbias$ spectra.
While all the peaks appear in the reverse $\Vbias$ sweep (positive to negative),
the first peak disappears in the forward sweep.
Figure~\ref{fig:switchers}d-g shows a more dramatic change.
During the scan, a sudden switch in the confinement potential occurs,
leading to the transition from the single QD (Fig.~\ref{fig:switchers}d)
to a coupled double QD (Fig.~\ref{fig:switchers}e).
This double QD state could be switched back to the single state 
(Fig.~\ref{fig:switchers}g) by scanning over the same area with a positive
$\Vbias$ (Fig.~\ref{fig:switchers}f).
Although both of these changes are readily identified in the images,  
having a spectrum alone may cause confusion
as is the case in conventional transport measurements.
These observations indicate that the QD confinement potentials are very sensitive
to the electrostatic background and can be modified, or switched, controllably.

Charge sensing with AFM can be used to investigate
the electronic structure 
of single and coupled self-assembled QDs.
The technique enables the quantitative extraction of the tunneling rate, 
charging energy, and the QD interaction energies; 
further, we have used it to perform the first measurement of 
temperature-dependent Coulomb blockade peak shifts confirming the 
shell degeneracy of the QD.
The dissipation images proved especially useful in analyzing multiple QDs and 
changes in QD confinement potential resulting from nearby charge fluctuations.
The images also revealed 
that what looked like a single QD structure topographically can actually contain multiple QDs.
Additionally, the imaging capability of AFM provides insight into the link
between QD electronic structure and 
topography which is of great importance 
in developing electronic devices based on QDs.

\section*{Methods}
The sample, grown by chemical beam epitaxy \cite{Poole01}, 
consists of the following layers:
460~nm undoped InP grown on top of an insulating InP substrate,
followed by a 10~nm Si-doped InP layer, 10~nm undoped layer, 
10~nm In$_{0.53}$Ga$_{0.47}$As layer, 20~nm undoped layer 
and a 1.82~ML InAs layer that results in the 
formation of InAs QDs by Stranski-Krastanow growth.
The QDs cover the surface with at a density of $\sim$ 2.5 QDs per $\mu \mathrm{m}^2$  
having diameters in the range of 30-95~nm and heights of 0.5-6~nm. 
The 2DEG layer formed in the InGaAs well serves as a back electrode and 
an Ohmic contact to the 2DEG is made by indium diffusion.

Our home-built cryogenic AFM \cite{Roseman00} includes an 
RF- modulated fiber optic interferometer
\cite{Rugar89} with 1550~nm wavelength for cantilever position detection.  
We coat Si AFM cantilevers (Nanosensors PPP-NCLR) with 10~nm titanium (adhesion layer) and 
20~nm platinum.  
The cantilevers typically have a 160~kHz resonance frequency and a 
quality factor between 100,000 and 200,000 at 4.5~K.

All of the images were taken in frequency modulation mode \cite{Albrecht91}.
In this mode, the cantilever is self-oscillated
at its resonance frequency
with a constant amplitude.
The frequency shift and dissipation were measured with a commercially available 
phase-locked loop frequency detector (Nanosurf, easyPLL plus).  
The topography images were taken 
in constant frequency shift mode 
where a constant frequency shift 
is maintained by regulating the cantilever tip-sample distance 
using a feedback controller.
The frequency shift and dissipation images were taken 
in constant-height mode with a typical tip height of 20~nm.
Dissipation images are shown in Fig.~S1 as a
function of $\Vbias$.
More negative $\Vbias$ results in adding more
electrons to the quantum dot.
Areas of increased dissipation mark 2DEG-QD tunneling events.
Each time a ring is crossed when traveling towards the quantum
dot center marks the addition of an electron to the dot.
More details of the AFM images are listed in Supplementary Table 1.

The amplitude of the cantilever excitation signal, $A_\mathrm{exc}$, 
is provided 
as the dissipation signal from the Nanosurf oscillator controller.
It is converted to units of 1/s via: 
$\frac{\omega_0}{Q} \frac{A_\mathrm{exc}-A_\mathrm{exc0}}{A_\mathrm{exc0}}$.
$A_\mathrm{exc0}$ is the excitation amplitude 
independent of the tunneling process, in other words the background 
dissipation.  
This conversion is independent of cantilever oscillation amplitude.  
Similarly, the signal is converted to units of eV/cycle 
by multiplying $\frac{A_\mathrm{exc}-A_\mathrm{exc0}}{A_\mathrm{exc0}}$ 
by the factor $E_0 = \frac{\pi k_0 a^2}{eQ}$ 
where $a$ is the cantilever oscillation amplitude \cite{Morita}.

The $\df-\Vbias$ spectra shown in Fig.~\ref{fig:spectra}a 
was originally superposed onto a large parabolic background 
arising from the capacitive force between the 2DEG and cantilever tip.  
Over several volts, at typical cantilever-sample gaps of 20~nm, 
the curve can be fit with a single parabola.  
In Fig.~\ref{fig:spectra}a this parabola was subtracted 
from the frequency shift data.

The exact positions of the peaks (dips) in the dissipation (frequency shift) 
are sensitive to the distance between cantilever tip and quantum dot.  
In particular, slight changes in cantilever tip lateral position 
with respect to the quantum dot center 
can lead to slight shifts in the peaks 
as can be deduced from the images 
where the rings can have different spacing depending on location.  
The shift in peaks as a function of height, however, is linearly dependent 
over the distances used in this experiment (12-22~nm).  
We took the differences in peak positions in the data displayed 
in Fig.~\ref{fig:spectra}b 
to be caused by small height differences (sub 1~nm) 
and so the voltage axis was rescaled to align the peaks with the data 
but the peak heights were not rescaled.  
The mean factor involved in the voltage rescaling is 1.011 
with the most extreme factor being 1.088.  
The temperature data above 22~K had 
thermally limited peaks 
for a cantilever oscillation amplitude of 0.4~nm, 
but needed to be reduced to 0.2~nm at 4.5~K.
The errorbars in Fig.~\ref{fig:spectra}c and d
represent how well the measurement over a single location was reproduced.

\section*{References}


\section*{Acknowledgement}
Funding for this research was provided by 
the Natural Sciences and Engineering Research Council of Canada, 
le Fonds Qu\'{e}b\'{e}cois de le Recherche sur la Nature et les Technologies, 
the Carl Reinhardt Fellowship,
and the Canadian Institute for Advanced Research.

\clearpage

\def\theequation{S\arabic{equation}}
\def\thefigure{S\arabic{figure}}

\setcounter{equation}{0}
\setcounter{figure}{0}

\begin{widetext}

  \begin{center}
    {\large \bf SUPPLEMENTARY INFORMATION}    
  \end{center}

\section{Figure Details}
The tip height for the voltage spectra and 
all of the constant height images was 19~nm $\pm$ 1~nm,
with the exception of Fig.~4d-g and Fig.~S1 
where the height was $\sim 23$~nm.
Additional image details are listed in Supplementary Table 1. 
The acquisition time of the majority of spectra was 15~seconds.

\begin{figure*}[h!]
  \includegraphics [width = 120mm] {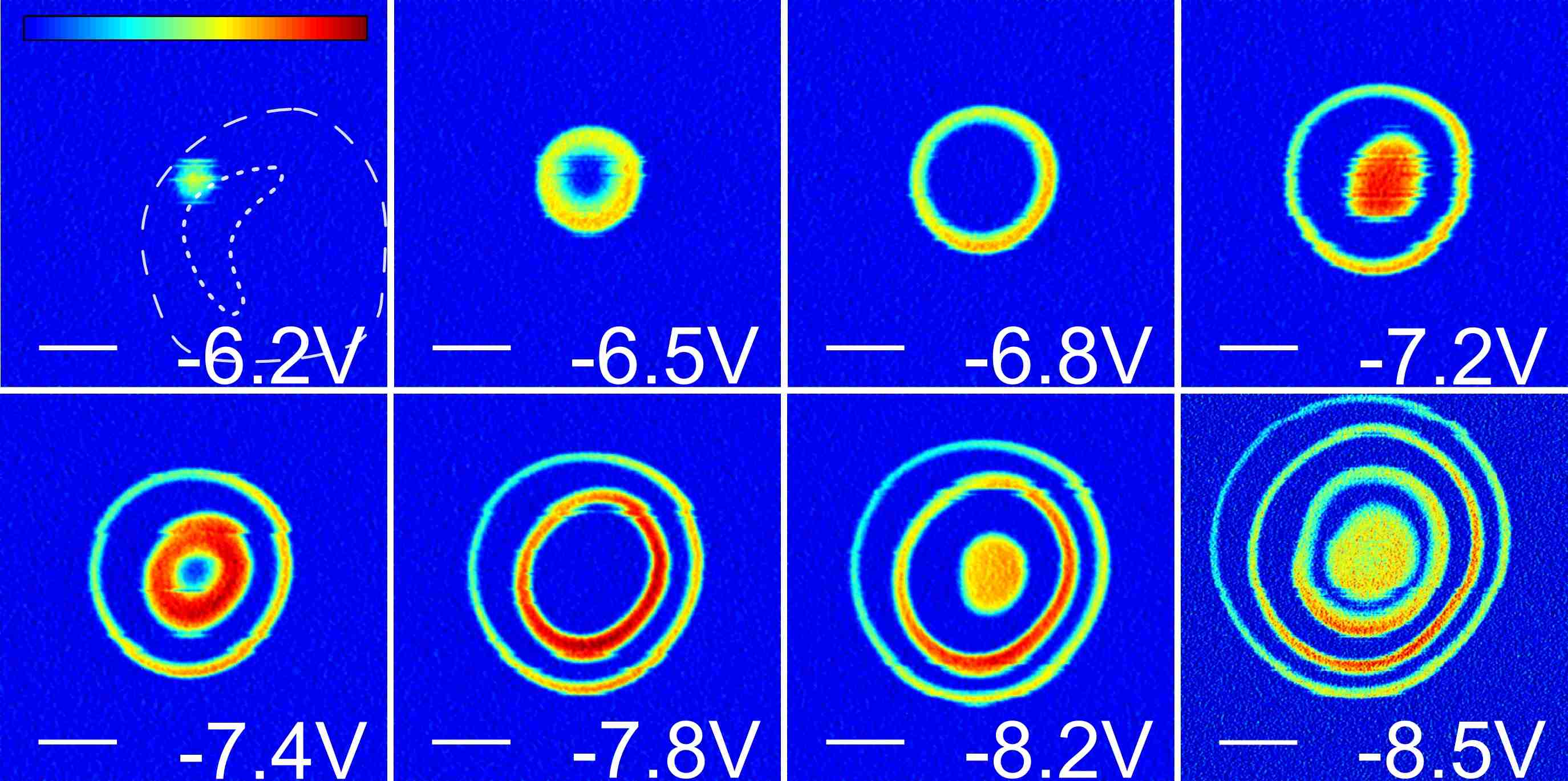}
  \caption{\label{fig:volts}
    A series of constant height dissipation images, for the 
    QD shown in Fig.~4d-g, for increasingly negative $\Vbias$.
    The base of the InAs structure is outlined with rectangular dashes and the highest area
    is outlined with rounded, more closely spaced, dashes.  
    This QD is localized near a high point in the structure, which is often observed.
    For increasingly negative $\Vbias$ more rings emerge as the QD is populated with electrons.  
    In these images the ring lineshape is broadened by the large cantilever
    oscillation amplitude of 0.33~nm at 4.5K with a tip-sample gap of roughly 23~nm.
    Note that the lateral position of the final image is slightly offset from the others as
    this image was taken at a later time in the experiment.
    Notice streaks appear in the same ring 
    location indicating some nearby electrostatic influence.
    The scalebar is 20~nm.  
    The same colorbar was used for each image, with all images but the last having
    a range of 0--0.85~Hz and the last 0--2~Hz.
  }
\end{figure*}

\begin{table*}[h!]
\caption{Experimental details of AFM images}
\begin{tabular}{l c c c c c} \hline
Fig. & T (K) & $\df/2\pi$ (Hz) & Oscillation Amplitude (nm) & $\Vbias$ (V) & Acquisition Time (min.)\\ \hline
1c & 78 & -9.4 & 1.6 & -0.35 & 6\\
1d & 4.5 & -- & 0.4 & -8.0 & 119\\
1e & 4.5 & -- & 0.4 & -8.0 & 119\\
3a & 4.5 & -- & 0.4 & -9.0 & 51\\
3b & 4.5 & -- & 0.4 & -7.6 & 17\\
3c & 78 & -9.4 & 1.6 & -0.35 & 14\\
3d-e & 4.5 & -- & 0.4 & -8.0 & 51\\
4a & 78 & -9.4 & 1.6 & -0.35 & 9\\
4b & 78 & -- & 1.6 & -8.0 & 68\\
4c & 4.5 & -- & 0.4 & -8.0 & 51\\
4d & 4.5 & -- & 0.8 & -8.0 & 9\\
4e & 4.5 & -- & 0.8 & -9.0 & 9\\
4f & 4.5 & -- & 0.8 & +6.8 & 9\\
4g & 4.5 & -- & 0.8 & -8.0 & 9\\
5  & 4.5 & -- & 0.8 & -- & 17 (last image 9)\\ \hline
\end{tabular}
\end{table*}

\section{Details of dissipation with degenerate shells}
Here we outline the approach used to derive
the general expression for the dissipation in
Eq.~(2).
The charging Hamiltonian for small cantilever tip motion 
may be written
\begin{align}
  \label{eq:Hc}
  H_{\mathrm{C}} &= 
  \sum_N  E_{\text{C}_N}
  \left[ \left( N - \Nvoltage \right)^2  
  + \left( 1 + \frac{C_\text{2DEG}}{C_\text{tip}}  \right) \Nvoltage^2
  \right]
  \ket{N} \bra{N}
  \\
  &= H_{\mathrm{C},0} + \Delta H_\text{osc} - \sum_N 
  A_N N z \ket{N} \bra{N}, \nonumber
\end{align}
where $\ket{N}$ is a state with $N$ electrons on the QD,
$\Nvoltage = - \frac{C_\text{tip} \Vbias}{e}$ is the dimensionless gate voltage,
$C_\text{tip}$ is the QD-tip capacitance and
$C_\text{2DEG}$ is the QD-2DEG capacitance.
In the second line,
$H_{\mathrm{C},0}$ is the cantilever-independent part of the
charging Hamiltonian,
$\Delta H_\text{osc}$ describes an electrostatic modification
of the cantilever potential,
and the QD-cantilever coupling strength for given $N$ is
$A_N = -2E_{\text{C}_N} \frac{\Vbias}{e} \left( 1-\alpha \right) 
\frac{\partial C_\mathrm{tip}}{\partial z}$.
We emphasize that 
$\Ec$ and $A$ may be different for each
electron added, as indicated by the index $N$.
From the second equality in equation (\ref{eq:Hc}),
$N$ plays the role of a force on the cantilever.
As a result, the dissipation and frequency shift
may be found from the linear response
coefficient $\lambda_N(\omega)$
describing the response of $N$ to changes in $z$ \cite{Clerk05}.

Consider the 
charge degeneracy point between $N$ and $N+1$
electrons on the QD.
This may always be viewed as
$\nShell$ or $\nShell+1$ electrons occupying a shell
of degeneracy $\nu$
(even for a nondegenerate single level, for which
$\nShell = 0$ and $\nu = 1$).
Neglecting interactions,
the charge state with $\nShell$ ($\nShell+1$) 
electrons
in the shell is $\Dn$-fold ($\Dnn$-fold) degenerate,
with
\begin{equation}
	\Dn = \binom{\nu}{\nShell}
	\quad,\quad
	\Dnn = \binom{\nu}{\nShell + 1},
\end{equation}
where $\binom{\cdot}{\cdot}$ denotes a binomial coefficient.
These arise simply from the different ways
to put $\nShell$
or $\nShell + 1$ electrons into
$\nu$ single particle states.
Let $P_{\nShell,i}$ be the probability to find $\nShell$ electrons
occupying the shell in configuration $i$,
and $P_{\nShell+1,j}$ be the probability to
find $\nShell+1$ electrons occupying the shell in
configuration $j$.
In general, these probabilities will satisfy 
the master equations
\cite{Beenakker91}
\begin{align}
	\partial_t P_{\nShell,i} &= \sum_j \left\{
	\Gamma_{j \rightarrow i}  P_{\nShell+1,j} 
	-\Gamma_{i \rightarrow j} P_{\nShell,i} 
	\right\}
	\label{eq:ME1} \\
	\partial_t P_{\nShell+1,j} &=
	 \sum_i \left\{
	\Gamma_{i \rightarrow j} P_{\nShell,i} 
	- \Gamma_{j \rightarrow i} P_{\nShell+1,j} \right\},
	\label{eq:ME2}
\end{align}
where $\Gamma_{i \rightarrow j}$  is the rate to
add an electron to configuration $i$ producing configuration
$j$, and vice versa for 
$\Gamma_{j \rightarrow i}$
(note that these rates are nonzero only for configurations
$i$ and $j$ that differ by the addition or removal
of one electron).
We calculate the rates using Fermi's golden rule.

The master equations (~\ref{eq:ME1}) and (~\ref{eq:ME2}) may be solved
for given values of $\nu$ and $\nShell$, but
the solutions are cumbersome for highly degenerate shells.
To simplify the equations we assume that 
for a given charge degeneracy point (i.e. a single dissipation peak), 
the tunneling
matrix elements from Fermi's golden rule are
equal for all single particle states
within the relevant shell.  
This is an approximation, since
degenerate states may indeed have different
wavefunctions leading to
different tunneling rates.
However, we expect the rates to be similar since
the tunnel barrier between the QD
and the 2DEG extends over the entire QD area, minimizing
the effects of the spatial variations of
different wavefunctions.
Moreover, we checked that 
significantly unequal rates lead only to very
small corrections in the peak shifts.
For example, taking distinct
rates for the two degenerate
orbital states in the $p$ shell,
we find that rates differing by a factor
of 2 lead to a correction
of 1.5\% for the shift
of the 3rd dissipation peak (i.e.~the 1st peak in the $p$ shell).
We thus neglect these possible differences here.
Taking the rates to be equal we arrive at the simplified
master equation for the total probability to find
$\nShell$ electrons in the shell,
\begin{align}
\label{eq:ME}
	\partial_t P_\nShell &= (\nu-\nShell)
	\left[
	\frac{\Dn}{\Dnn}  \Gm (1- P_\nShell) - \Gp P_\nShell
	\right]
\end{align}
where
\begin{align}
	\Gamma_+ = \Gamma f(E)
	\quad,\quad
	\Gamma_- = \Gamma \left[1- f(E)\right]
\end{align}
are the rates to add (+) or remove 
($-$)
an electron to or from a single particle state,
and $f$ is the Fermi function.
Note that the master equation for $P_{\nShell+1}$
is not independent in our approximation of equal rates;
this is a result of
$P_\nShell + P_{\nShell+1} = 1$.
The stationary solution of equation (\ref{eq:ME}) is
\begin{align}
\label{eq:pStat}
	P_\nShell &= \frac{(\nShell+1)}{\phi}(1-f) \\
	P_{\nShell+1} &= \frac{(\nu-\nShell)}{\phi} f,
\end{align}
where $\phi$ is defined in equation~(3).

The quantity we need is
the linear response coefficient
$\lambda_N(\omega)$.
To find this, we assume that the cantilever is
oscillating at frequency $\omega$.
This causes the
chemical potential difference between
the QD and the 2DEG
to oscillate,
\begin{equation}
\label{eq:Eosc}
	E \rightarrow E + \delta e^{-i\omega t},
\end{equation}
and this leads to a change in the probabilities,
\begin{align}
	P_{\nShell+1} &\rightarrow  P_{\nShell+1} + \lambda_N(\omega) \delta e^{-i\omega t}, \\
	P_{\nShell} &\rightarrow P_\nShell - \lambda_N(\omega) \delta e^{-i\omega t}.
	\label{eq:pChange}
\end{align}
Inserting equations (\ref{eq:pStat})--(\ref{eq:pChange}) 
into equation (\ref{eq:ME}) and linearizing in $\delta$, we solve
 for $\lambda_N(\omega)$.
Its real and imaginary part yield 
the dissipative and 
conservative parts of the electrostatic force from
$(k_0/\omega_0^2) \gamma = - A^2 \Im{\left\{ \lambda_N(\omega) \right\}} / \omega $
and $(2k_0/\omega_0) \Delta \omega = A^2 \Re{\left\{ \lambda_N(\omega)\right\}}$.
The dissipation for arbitrary degeneracy is given in equation~(2).
and for the frequency shift we obtain
\begin{equation}
\label{eq:freqShiftGen}
	\Delta \omega = -\frac{\omega_0}{2k_0} \frac{A^2 \Gamma^2}{\kBol T}
	\left[
	\frac{(\nShell+1)(\nu-\nShell)}{\omega^2 + (\phi \Gamma)^2} 
	\right]
	f(1-f)	.
\end{equation}
Note that we recover the single level result
[i.e. equation~(1) for the dissipation]
by taking $\nu = 1$ and $\nShell = 0$ as expected.
Finally, we point out that the temperature-dependent
level repulsion discussed in the paper is contained in a symmetry
of equations~(2) and (\ref{eq:freqShiftGen}), from which
we find that taking $\nShell \rightarrow \nu-\nShell -1$
is equivalent to $E \rightarrow -E$.

The peak shifts of $\gamma$ and $\Delta \omega$ are
proportional to temperature and we can solve for the coefficients
analytically.
However, in general the coefficients are complicated and
unenlightening.
To show how the peak shifts depend on degeneracy,
we provide the coefficients in the low and high frequency limits
where they are greatly simplified.  Note that our experiment is in the intermediate
regime $\omega \sim \Gamma$, so the
peak shifts measured and calculated
in the main text lie between these two limits.
For $\gamma$, the peak shifts in the low and high frequency limits are
\begin{equation}
	\frac{\Delta E_{\gamma,\text{peak}}}{\kBol T} \rightarrow
	\begin{cases}
	\ln{\left( d + \sqrt{d(d+1) +1} \right)} &\text{as } (\omega\rightarrow 0), \\
	\ln{ \sqrt{d+1} } &\text{as } (\omega\rightarrow \infty),
	\end{cases}
\end{equation}
where
\begin{equation}
	d = \frac{\nu-\nShell}{\nShell+1} - 1.
\end{equation}
For a nondegenerate level, $d=0$ and there is no peak shift
at any frequency.
For $\Delta \omega$ the peak shifts in the same two limits are
\begin{equation}
	\frac{\Delta E_{\Delta\omega,\text{peak}}}{\kBol T} \rightarrow
	\begin{cases}
	\ln{\left( d + 1\right)} &\text{as } (\omega\rightarrow 0), \\
	0 &\text{as } (\omega\rightarrow \infty).
	\end{cases}
\end{equation}
Comparing these limits, we see that
the shell degeneracy results in a greater
peak shift in
$\gamma$ than in $\Delta \omega$.
This is a direct consequence of equation~(4),
from which we see that, aside from an energy-independent
prefactor,
$\Delta\omega$ differs
from $\gamma$ by a factor of $\phi$.

We measured the separation between the peak in $\gamma$
and the peak in $\Delta \omega$ for each charge degeneracy point.
This is shown in Fig.~\ref{fig:peak3} for the third peak in
Fig.~2a as a function of temperature and compared to theory
with no fit parameters.
As argued in the main text, this provides strong evidence
that the observed peak shifts are indeed a
result of shell degeneracy.


\begin{figure*}
\includegraphics [width = 100mm] {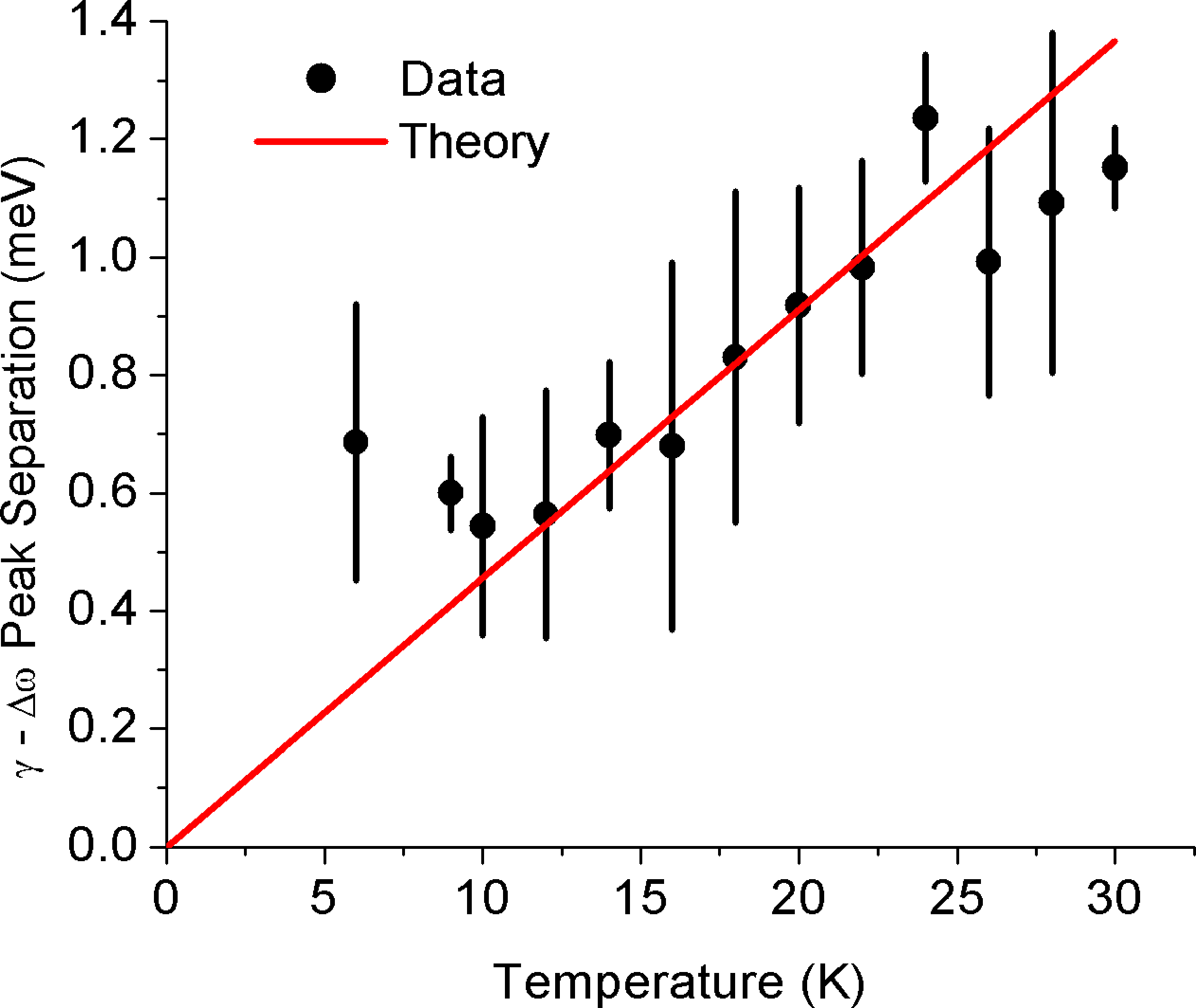}
\caption{\label{fig:peak3}
Difference between dissipation and frequency shift peak
positions as a function of temperature for peak 3, compared to 
theoretical prediction with no fit parameters.
}
\end{figure*}

\end{widetext}

\begin{thebibliography}{34}
\bibitem[{Ste()}]{StevesFootnote}
\bibinfo{note}{We checked that taking distinct rates for each degenerate state
  leads to no qualitative change in the results.}
\end{thebibliography}


\end{document}